\documentclass[]{raa}
\usepackage{graphicx,times}
\usepackage{natbib}
\usepackage{indentfirst}
\usepackage{ifpdf}
\usepackage{footnote}

\providecommand{\bm}[1]{\mbox{\boldmath$#1$\unboldmath}}

\begin{document}

   \title{The beaming effect and $\rm{\gamma}$-ray emission for Fermi blazars
$^*$}

 \volnopage{ {\bf 2015} Vol.\ {\bf X} No. {\bf XX}, 000--000}
   \setcounter{page}{1}

   \author{Yong-Yun Chen\inst{1}, Xiong Zhang$^{\dag}$\inst{2}, Dingrong Xiong\inst{3}
   ,Si-Ju Wang\inst{4},Xiaoling Yu\inst{5}}

   \institute{Department of Physics, Yunnan Normal University, Kunming 650500,
China \\
        \and
            Department of Physics, Yunnan Normal University, Kunming 650500,
China\\$^{\dag}$e-mail:ynzx@yeah,net.
         \and
            National Astronomical Observatories/Yunnan Observatories, Chinese Academy of Sciences, Kunming 650011,
China \\
          \and
          Dongxing Middle School, Chuxiong,Yunnan 675000,
China \\
           \and
      Department of Physics, Yunnan Normal University, Kunming 650500,
China \\
{\small Received ********; accepted ********}}

\abstract{We study the $\rm{\gamma}$-ray luminosity and beaming effect for Fermi blazars. Our results are as follows. (i) There are significant correlations between $\rm{\gamma}$-ray luminosity and radio core luminosity, and between $\rm{\gamma}$-ray luminosity and $\rm{R_{v}}$, which suggests that the $\rm{\gamma}$-ray emissions have strong beaming effect. (ii) Using the $\rm{L_{ext}/M_{abs}}$ as an indicator of environment effects, we find that there have no significant correlation between $\rm{\gamma}$-ray luminosity and $\rm{L_{ext}/M_{abs}}$ for all sources when remove the effect of redshift. FSRQs considered alone also do not show a significant correlation, while BL Lacs still show a significant correlation when remove the effect of redshift. These results suggest that the $\rm{\gamma}$-ray emission may be affected by environment on the kiloparsec-scale for BL Lacs.
}

   \authorrunning{Y.Y.Chen}            
   \titlerunning{beaming effect and $\rm{\gamma}$-ray emission}  
   \maketitle

%
\section{Introduction}
Blazars are an extreme subclass of AGNs. According to the observations, blazars show some extreme properties, namely, rapid variability, high optical polarization, high-energy $\rm{\gamma}$-ray emission and superluminal motion (Angle \& Stockman 1980; Wills et al. 1992; Stickel et al. 1993; Andruchow et al.2005; Fan 2005; Gu et al.2006). Blazars are often divided into BL Lacertae objects (BL Lacs) and flat spectrum radio quasars (FSRQs) based on their emission line features. The FSRQs have strong emission lines, while BL Lacs have only very weak or non-existent emission lines. The classical divisions between FSRQs and BL Lacs are mainly based on the equivalent width (EW) of the emission lines. Those blazars with rest frame EW$<5\rm{\AA}$ are classified as BL Lacs (e.g., Urry \& Padovani 1995). Many authors have found that there are many similarities characteristics between FSRQs and BL Lacs, therefore most authors thought that they should be regarded as a single class which have blazars behavior (Angel \& Stockman 1980; Ghisellini et al.2011; Giommi et al.2012). However, some opponents argue that FSRQs and BL Lacs should not be treated as a single category due to the different emission lines between them. Ghisellini et al. (2009a) found that FSRQs and BL Lacs are well separated into two populations in the $\rm{\alpha_{\gamma}-L_{\gamma}}$ plane, where $\rm{\alpha_{\gamma}}$ and $\rm{L_{\gamma}}$ are the spectral index and $\rm{\gamma}$-ray luminosity. Ackermann et al. (2011) have also confirmed this result.

Since the launch of the Fermi satellite, we have entered in a new era of blazars research (Abdo et al. 2009, 2010). Up to now, the Large Area Telescope (LAT) has detected hundreds of blazars because it has high sensitivity than EGRET in the 0.1-100 Gev energy rang. However, the questions why some sources are $\rm{\gamma}$-ray loud and others are gamma-ray quiet are still unclear. Doppler boosting is believed to be one of the important answers to the questions. Blazars detected by LAT are more likely to have larger Doppler factors (e.g., Lister et al. 2009; Savolainen et al. 2010). Linford et al. (2011) found that the difference between the $\rm{\gamma}$-ray loud and quiet FSRQs can be explained by Doppler boosting. The vast majority of the LAT $\rm{\gamma}$-ray sources are blazars, with strong, compact radio emission. These blazars show flat radio spectral, compact cores with one-sided parsec-scale jets (Marscher 2006). Many authors suggest that the $\rm{\gamma}$-ray emission originates from the jet and is also relativistically beamed, in the same manner as the radio emission (e.g.,von Montigny et al. 1995; Mattox et al. 1993). The many correlations found between the $\rm{\gamma}$-ray emissions detected by EGRET and the radio/mm-wave properties of blazars further support this scenario (Valtaoja \& Ter$\rm{\ddot{a}}$sranta 1995; Jorstad et al. 2001a,b; L$\rm{\ddot{a}}$hteenm$\rm{\ddot{a}}$ki \& Valtaoja 2003; Kellermann et al. 2004; Kovalev et al. 2005).

In this paper, we study the beaming effect and $\rm{\gamma}$-ray emissions for Fermi blazars. The paper is structured as follows: we present the sample in Sect.2; the results and discussions are in Sect.3; conclusions are in Sect.4.
\section{Sample and data description}
A large number of blazars that have gamma-ray emission are being detected by the Fermi Large Area Telescope (LAT). The second LAT catalog (2LAC) containe 886 clean sample, comprising 395 BL Lacs, 310 FSRQs, 157 candidate blazars of unknow types, 8 misaligned AGNs, 4 narrow line Seyfert1, 10 AGNs of other types and 2 starbust galaxies (Ackermann et al.2011; Abdo et al.2012). The recent release is the third Fermi catalog of $\rm{\gamma}$-ray sources, based on 4 years of data, and consisting of 3033 sources (The Fermi-LAT Collaboration, 2015).

We tried to select the large number blazars with reliable redshift, radio core and extended radio luminosity at 1.4 GHz. Firstly, we considered the following samples of blazars to get the radio core luminosity and extended luminoaity at 1.4 GHz: Kharb et al. (2010), Antonucci \& Ulvestad (1985), Cassaro et al. (1999), Murphy et al. (1993), Landt et al. (2008), Caccianiga et al. (2004), Giroletti et al.(2004). We also got the radio core luminosity at 1.4 GHz from the NED for few sources, which do not mark the references in Table 1. Secondly, cross-correlating these sample with the Fermi LAT Third Sources Catalog (3FGL), and we got the 3FGL spectral index and energy flux at 0.1-100 Gev from clean Fermi LAT of Third Source Catalog (3FGL, Fermi-LAT Collaboration 2015)\footnote{The 3FGL catalog is available at http://fermi.gsfc.nasa.gov/ssc/data/access/lat/4yr$\_$catalog}. At the same time, these sources are also included in the clean Second Catalog (2FGL). At last, we got 201 Fermi blazars. We used $\rm{L_{v}=4\pi d_{L}^{2}S_{v}}$ to calculate the luminosity, the flux was k-corrected by $\rm{S_{v}=S_{v}^{obs}(1+z)^{\alpha-1}}$, $\rm{\alpha}$ is spectral indexs, $\rm{\alpha^{\gamma}=\alpha^{ph}-1}$, $\alpha^{ph}$ is photon spectral index. The $\rm{\gamma}$-ray luminosity can be calculated by the above equations.

The relevant data for Fermi blazars is listed in Table 1 with the following headings: column (1) the third name of sources (3FGL); column (2) the classification of sources (BZB=BL Lac objects, BZQ=flat-spectrum radio quasars); column (3) redshift; column (4) apparent V-band magnitude obtained from the AGN catalog of V$\rm{\acute{e}}$ron-Cetty \& V$\rm{\acute{e}}$ron (2006); column (5) the 3FGL photo-spectral index; column (6) the 3FGL gamma-ray energy flux in the 0.1-100 Gev; column (7) the radio core flux at 1.4 GHz, the units is Jy; column (8) the extended radio flux at 1.4 GHz, the units is mJy; column (9) the references of column (7) and (8).

\section{Results and discussions}
\subsection{The distributions of energy spectral index and $\rm{\gamma}$-ray luminosity}
Figure 1 shows the energy spectral index as a function of redshift (left) and $\rm{\gamma}$-ray luminosity as a function of redshift (right). We find that the energy spectral indexes do not have significant dependence on redshift if blazar subclasses are considered separately. Ackermann et al.(2011) also found that there is no significant dependence of the photon index on redshift is observed if blazars subclasses are considered separately by using the blazars in 2LAC, as illustrated in Figure 19 of Ackermann et al.(2011). Our result contrary to the result of Ackermann et al.(2011). We find that there are significant correlation between spectral index and redshift for BL Lacs by using 3LAC (see Table2). The $\rm{\gamma}$-ray luminosity is plotted as a function of redsift in the right of Figure 1. A Malmquist bias is readily apparent in this figure as only high $\rm{\gamma}$-ray luminosity sources (mostly FSRQs) are detected at large distances. Given their $\rm{\gamma}$-ray luminosity distribution, most BL Lac objects could not be detected if they were located at redshifts, which are greater than 1.

Figure 2 shows the $\rm{\gamma}$-ray energy index versus $\rm{\gamma}$-ray luminosity plane. From Figure 2, we can find that the spectral index vs $\rm{\gamma}$-ray luminosity plane reveal a clean separation between FSRQs and BL Lacs. This correlation has been discussed in detail in the context of the ``blazar divide'' (Ghisellini et al. 2009a). We find that most FSRQs have $\rm{L_{\gamma}>10^{46} erg s^{-1}}$, while BL Lacs are below this value. Ghisellini et al. (2009a) suggested that it is interpreted as a consequence of the changing accretion regime of the underlying accretion disc from radiatively efficient to inefficient, or, in other words, from a standard Shakura \& Sunyaev (1973) disc to an advection dominated accretion flow (ADAF). We also find that most BL Lacs have a relative flat $\rm{\alpha_{\gamma}}$ ($\rm{\alpha_{\gamma}<1.2}$).
\begin{figure}
   \centering
   \includegraphics[width=14.0cm, angle=0]{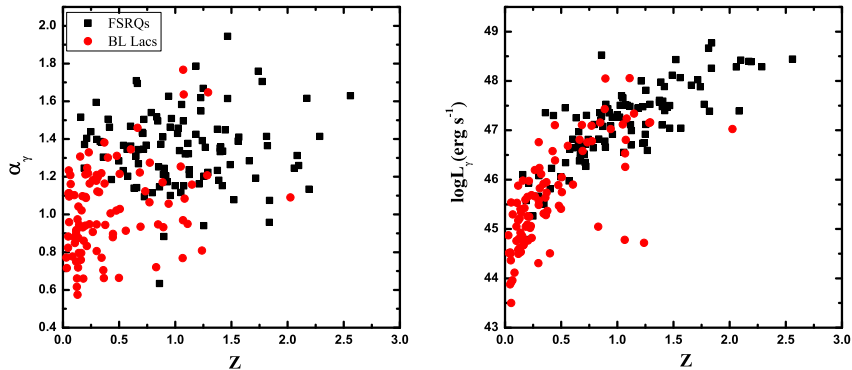}
   \caption{The Energy spectral index (left) and $\rm{\gamma}$-ray luminosity (right) vs redshift for Fermi blazars. FSRQs: the black squares; BL Lacs: filled red circle. }
   \label{Fig1}
   \end{figure}
\begin{figure}
   \centering
  \includegraphics[width=9cm, height=8.6cm]{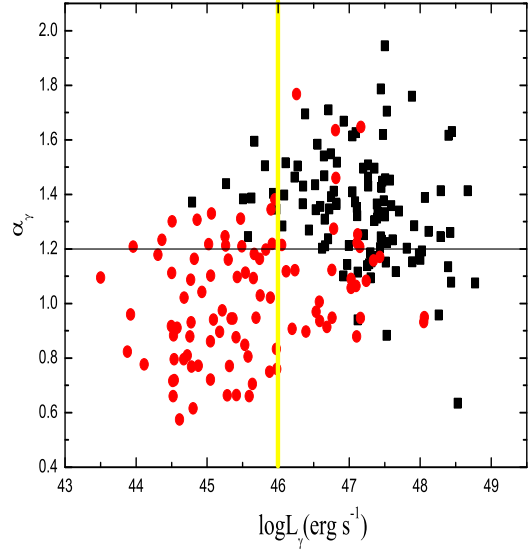}
   \caption{Energy spectral index vs $\rm{\gamma}$-ray luminosity for Fermi blazars. The black line is $\rm{\alpha_{\gamma}=1.2}$; the yellow line is $\rm{L_{\gamma}=10^{46}erg s^{-1}}$. The meanings of different symbols are as same Fig.1.}
   \label{Fig2}
   \end{figure}
\subsection{The $\rm{\gamma}$-ray luminosity vs radio core and extended luminosity}
Figure 3 shows the $\rm{\gamma}$-ray luminosity as a function of radio core luminosity (left) and $\rm{\gamma}$-ray luminosity as a function of radio extended luminosity (right). We find a significant correlation between $\rm{\gamma}$-ray luminosity and radio core luminosity for our sample (Table 2). Partial regression analysis also shows that the linear correlation between $\rm{\gamma}$-ray luminosity and radio core luminosity is significant when the effects of redshift are removed ($\rm{r_{XY.Z}}$=0.766, $\rm{P=7.13\times10^{-33}}$). This result suggest that $\rm{\gamma}$-ray have strong beaming effect. Linford et al. (2011) found that $\rm{\gamma}$-ray loud and $\rm{\gamma}$-quiet objects are related to the core, and they suspected that the $\rm{\gamma}$-ray radiation originates within the core (i.e., at the base of the jet). There is also a significant correlation between $\rm{\gamma}$-ray luminosity and extend radio luminosity (Table 2). Partial regression analysis also shows that the linear correlation between $\rm{\gamma}$-ray luminosity and radio extended luminosity is significant correlation when the effects of redshift are removed ($\rm{r_{XY.Z}}$=0.644, $\rm{P=7.31\times10^{-18}}$). Fan et al. (2014) suggested that the $\rm{\gamma}$-ray emissions are composed of two components, one is beamed, the other is unbeamed. The unbeamed part should be associated with the extend radio emissions. Recently, $\rm{\gamma}$-ray emissions were detected from the lobe of Cen A (Massaro \& Ajello 2011). Further investigation should be interesting.
\begin{figure}
   \centering
  \includegraphics[width=14.5cm, angle=0]{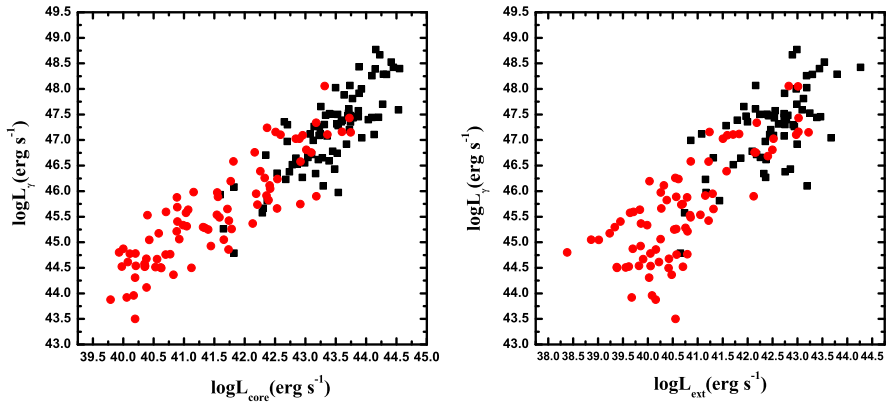}
   \caption{The $\rm{\gamma}$-ray luminosity vs 1.4 GHz core (left) and extended (right) luminosity for Fermi blazars. The meanings of different symbols are as same Fig.1.}
   \label{Fig3}
   \end{figure}
\begin{figure}
   \centering
  \includegraphics[width=14.5cm, angle=0]{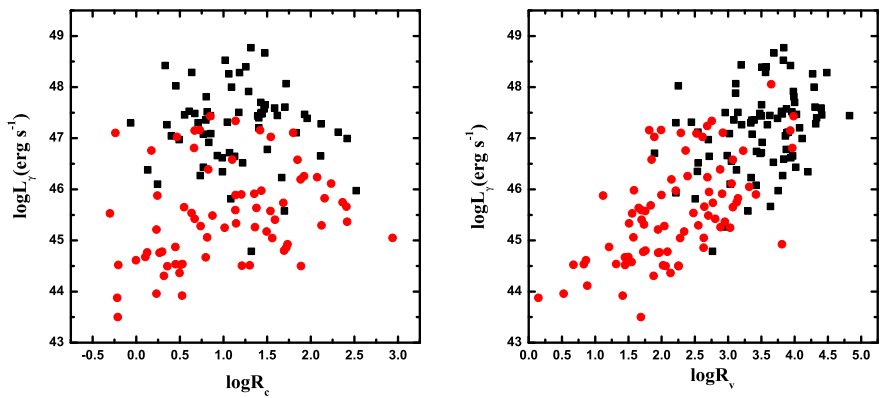}
   \caption{The $\rm{\gamma}$-ray luminosity vs orientation indicators, $\rm{R_{c}}$ (left) and $\rm{R_{v}}$ (right) for Fermi blazars. The meanings of different symbols are as same Fig.1.}
   \label{Fig4}
   \end{figure}
\subsection{The beaming effect}
The ratio of the beamed radio core flux density ($\rm{S_{core}}$) and the unbeamed extend radio flux density ($\rm{S_{ext}}$), namely, the radio core-dominance parameter ($\rm{R_{c}}$), has routinely been used as a statistical indicator of Doppler beaming and thereby orientation (Orr \& Browne 1982; Kapahi \& Saikia 1982; Kharb \& Shastri 2004). The k-corrected $\rm{R_{c}}$ ($\rm{=\frac{S_{core}}{S_{ext}} (1+z)^{\alpha_{core}-\alpha_{ext}}}$,with $\rm{\alpha_{core}=0}$, $\rm{\alpha_{ext}=0.8}$) is plotted against the $\rm{\gamma}$-ray luminosity in Figure 4 (left). As expected, $\rm{R_{c}}$ is correlated with the $\rm{\gamma}$-ray luminosity. However, we find that there is no significant correlation between $\rm{R_{c}}$ and $\rm{\gamma}$-ray luminosity (Table 2). Fan et al.(2014) also found that there have no significant correlation between $\rm{\gamma}$-ray luminosity and $\rm{R_{c}}$. But as we see below, the alternate orientation indicator, $\rm{R_{v}}$, dose show the expected behavior.
\begin{figure}
   \centering
  \includegraphics[width=9cm, height=8.6cm]{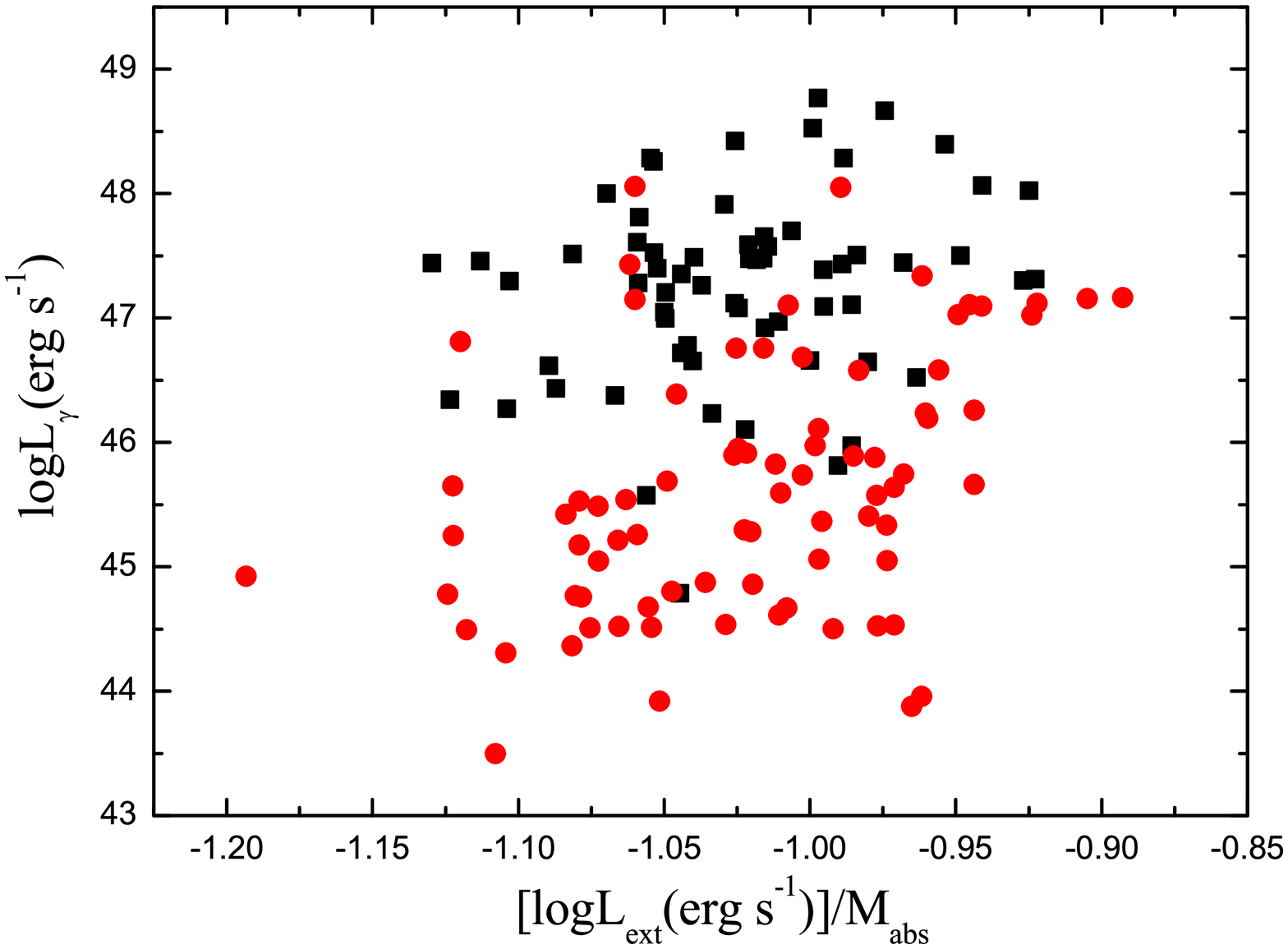}
   \caption{The $\rm{\gamma}$-ray luminosity vs the environment indicator for Fermi blazars. The meanings of different symbols are as same Fig.1.}
   \label{Fig5}
   \end{figure}
\begin{figure}
   \centering
  \includegraphics[width=9cm, height=8.6cm]{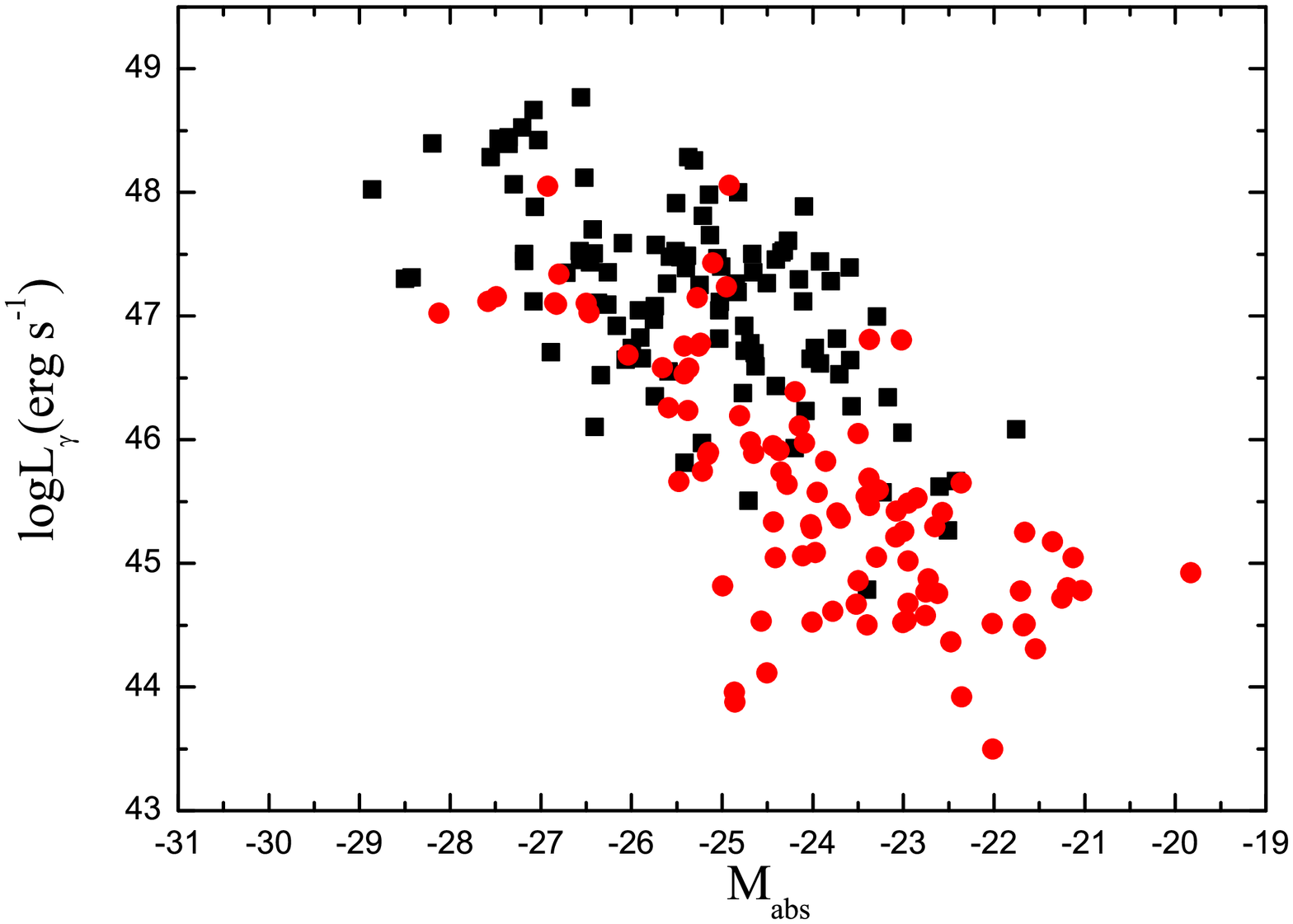}
   \caption{The $\rm{\gamma}$-ray luminosity vs absolute optical magnitude for Fermi blazars. The meanings of different symbols are as same Fig.1.}
   \label{Fig6}
   \end{figure}

Wills \& Brotherton (1995) defined $\rm{R_{v}}$ as the ratio of the radio core luminosity to the k-corrected absolute V-band magnitude ($\rm{M_{abs}}$):$\rm{\log{R_{v}}}$=$\rm{\log{\frac{L_{core}}{L_{opt}}}}$=$\rm{(\log{L_{core}}+M_{abs}/2.5)-13.7}$, where $\rm{M_{abs}=M_{V}-k}$ and the k-correction is $\rm{k}$=-2.5$\rm{\log}\rm{(1+z)^{1-\alpha_{opt}}}$ with the optical spectral index, $\rm{\alpha_{opt}=0.5}$. Kharb et al. (2010) found that the ratio of the radio core luminosity to the k-corrected absolute V-band magnitude ($\rm{R_{v}}$) is a better orientation indicator than $\rm{R_{c}}$ since the optical luminosity is likely to be a better measured of intrinsic jet power than extended radio luminosity (e.g., Maraschi et al. 2008; Ghisellini et al. 2009). This is due to the fact that the optical continuum luminosity is correlated with the emission line luminosity over 4 orders of magnitude (Yee \& Oke 1978), and the emission line luminosity is tightly correlated with the total jet kinetic power (Rawlings \& Saunders 1991). The extended radio luminosity is suggested to be affected by interaction with the environment on kiloparsec-scales. By making use of the above Equations, we obtain $\rm{R_{v}}$ for our Fermi blazars sample.  The $\rm{\gamma}$-ray luminosity have a strong beaming effect. Figure 4 suggests that $\rm{R_{v}}$ is indeed a better indicator of orientation as the correlation with $\rm{\gamma}$-ray luminosity gains in prominence (see Table 2). Xiong et al.(2015) have suggested that the Fermi and non-Fermi blazars have a significant difference in $\rm{R_{v}}$. However, there have no significant difference in $\rm{R_{c}}$ for Fermi and non-Fermi blazars. These results suggested that the $\rm{R_{v}}$ is a better orientation indicator than $\rm{R_{c}}$.
\subsection{The $\rm{\gamma}$-ray luminosity and environment effect}
Kharb et al. (2010) suggested that if the extend radio luminosity is indeed affected by interaction with the kiloparsec-scale environment and the optical luminosity is close AGN power, the ratio, $\rm{[\log{L_{ext}}]/M_{abs}}$, can serve as a probe for environment effects on kiloparsec-scales (as suggested by Wills \& Brotherton 1995). They used this ratio as the ``environment indicator''. According to the above equation, we obtain this ``environment indicator''.
Figure 5 shows the relation between $\rm{\gamma}$-ray luminosity and $\rm{[\log{L_{ext}}]/M_{abs}}$. We find that there have no significant correlation between them for all sources when remove the effect of redshift (Table 2). The FSRQs considered alone also do not show a significant correlation, while the BL Lacs still shows a significant correlation when remove the effect of redshift (Table 2).

Ghisellini et al.(2010) have suggested that the different properties between FSRQs and BL Lacs can be explained with the difference in jet power accompanied by a different environment, in turn caused by a different regime of accretion. FSRQs occur in the earlier phase. They have powerful disk and jet, high accretion. BL Lacs have weak disk and weaker lines emitted closer to the black hole. Dense environments can decrease expansion losses in the source, but increase radiative losses, making the sources brighter at low radio frequencies (Barthel \& Arnaud 1996). Because the around of BL Lacs have not dense environment, it may lead to the low accretion. The low accretion may lead to the low jet power. Ghisellini et al.(2010) have suggested that the $\rm{\gamma}$-ray luminosity is a good tracer of the jet power for blazars. They suggested that the jets of BL Lac objects thus propagate in a medium starved of external radiation (weak disk, weak lines), and this makes the emitting electrons accelerated in the jet to cool less, and to reach very high energies. Their emitted high energy spectrum, produced mainly by the synchrotron self-Compton mechanism,is less luminous and harder than in FSRQs. The latter sources in fact have a radiatively efficient accretion disk and a corresponding standard broad line region. If the jet dissipates most of its power within the broad line region, then the emitting electrons will efficiently cool (mostly by external Compton), will reach only moderate energies, and will produce a high energy peak below 100 MeV. Generally, the BL Lacs have lower $\rm{\gamma}$-ray luminosity than FSRQs. Zhang et al.(2014) have suggested that the dominating formation mechanism of FSRQ jets may be the Blandford-Znajek process, but BL Lac object jets may be produced via the Blandford-Payne and/or Blandford-Znajek processes, depending on the structures and accretion rates of accretion disks. The BP mechanism may power a jet by releasing the gravitational energy of accreting matter that moves toward the BH. The rotational energy of a rapidly rotating BH is essential for the BZ process. Ineson et al.(2015) have suggested that jet and environment have strong correlations for low accretion radio loud AGNs but not for High accretion radio loud AGNs. We know that the FSRQs are high accretion radio loud AGNs, and the BL Lacs are low accretion radio loud AGNs. Moreover, we find that there are significant correlation between $\rm{\gamma}-ray$ luminosity and environment indicator for BL Lacs, and not for FSRQs. These results therefore may suggest that the $\rm{\gamma}$-ray emissions are affected by the environment on kiloparsec-scales for BL Lacs.

Figure 6 shows the relation between $\rm{\gamma}$-ray luminosity and absolute magnitude. We find a significant anti-correlation between them (Table 2). This figure shows a similar ``blazar sequence''. Fossati et al. (1998) and Ghisellini et al. (1998) have proposed the so-called ``blazar sequence'' with plots of various powers vs the synchrotron peak frequency for a sample of blazars containing FSRQs and BL Lacs, of which an anti-correlation was apparent with the most powerful sources having relatively small synchrotron peak frequencies and the least powerful ones having the highest $\rm{\nu_{peak}}$ (Wu et al. 2008).

\section{Conclusions}
From our results and discussions mentioned above, we can conclude that (i) There are significant correlations between $\rm{\gamma}$-ray luminosity and both radio core luminosity and $\rm{R_{v}}$, which suggests that the $\rm{\gamma}$-ray luminosity have strong beaming effect. (ii) Using the $\rm{L_{ext}/M_{abs}}$ as a indicator of environment effects, FSRQs considered alone do not show a significant correlation, while BL Lacs still show a significant correlation when remove the effect of redshift. These results suggest that the $\rm{\gamma}$-ray emission may be affected by environment on the kiloparsec-scale for BL Lacs.
\normalem
\begin{acknowledgements}
We thank the anonymous referee for valuable
comments and suggestions. We are very grateful to the Science
Foundation of Yunnan Province of China(2012F13140,2010CD046). This work is supported by
the National Nature Science Foundation of China (11063004,11163007,U1231203), and the High-Energy Astrophysics Science and Technology Innovation Team of Yunnan Higher School and Yunnan Gravitation Theory Innovation Team (2011c1). This research has made use of the NASA/IPAC Extragalactic Database (NED), that is operated by Jet Propulsion Laboratory, California Institute of Technology, under contract with the National Aeronautics and
Space Administration.
\end{acknowledgements}

\clearpage
\begin{table*}
\centering
\begin{minipage}{105mm}
\caption{The sample of Fermi blazars.}
\begin{tabular}{@{}crcccccccccccccccrl@{}}
\hline\hline
3FGL name & Class & Redshift & V & $\rm{\alpha_{\gamma}}$ & $\rm{F_{\gamma} (erg s^{-1}cm^{-2})}$ & $\rm{S_{core}}$ & $\rm{S_{ext}}$ & Ref\\
{(1)} & {(2)} & {(3)} & {(4)} & {(5)} & {(6)} & {(7)} & {(8)} & {(9)}\\
\hline
3FGL J0050.6-0929	&	BZB	&	0.2	&	17.44	&	2.0931	&	4.03071E-11	&	0.57	&	139.7	&	K10	\\
3FGL J0108.7+0134	&	BZQ	&	2.099	&	18.39	&	2.260085	&	6.3754E-11	&	2.81	&	530.6	&	K10	\\
3FGL J0112.1+2245	&	BZB	&	0.265	&	15.66	&	1.907256	&	7.77878E-11	&	0.36	&	3.9	&	K10	\\
3FGL J0112.8+3207	&	BZQ	&	0.603	&	18.71	&	2.3626	&	3.87314E-11	&		&		&		\\
3FGL J0116.0-1134	&	BZQ	&	0.67	&	19	&	2.343904	&	1.51119E-11	&	0.65	&		&		\\
3FGL J0120.4-2700	&	BZB	&	0.559	&	16.21	&	1.914129	&	4.23886E-11	&		&	168	&	CB99	\\
3FGL J0132.6-1655	&	BZQ	&	1.02	&	19.21	&	2.430135	&	2.61622E-11	&		&		&		\\
3FGL J0137.0+4752	&	BZQ	&	0.859	&	19.5	&	2.146481	&	5.10975E-11	&	1.88	&	8.7	&	K10	\\
3FGL J0137.6-2430	&	BZQ	&	0.835	&	17.33	&	2.517637	&	1.53353E-11	&		&		&		\\
3FGL J0141.4-0929	&	BZB	&	0.733	&	17.5	&	2.122884	&	2.30759E-11	&		&	50	&	CB99	\\
3FGL J0205.2-1700	&	BZQ	&	1.739	&	17.92	&	2.759468	&	1.82214E-11	&	0.61	&		&		\\
3FGL J0204.8+3212	&	BZQ	&	1.466	&	17.4	&	2.944624	&	1.05638E-11	&	0.65	&	11.7	&	K10	\\
3FGL J0217.8+0143	&	BZQ	&	1.715	&	16.09	&	2.192922	&	4.61606E-11	&	0.45	&	71.1	&	K10	\\
3FGL J0222.6+4301	&	BZB	&	0.444	&	15.2	&	1.879783	&	1.93512E-10	&	0.814	&	1052	&	A85	\\
3FGL J0237.9+2848	&	BZQ	&	1.213	&	19.3	&	2.162098	&	1.11298E-10	&	2.33	&	99.9	&	K10	\\
3FGL J0238.6+1636	&	BZB	&	0.94	&	15.5	&	2.057468	&	1.00162E-10	&	1.51	&	25.5	&	K10	\\
3FGL J0252.8-2218	&	BZQ	&	1.419	&	20.41	&	2.152822	&	5.70612E-11	&		&		&		\\
3FGL J0259.5+0746	&	BZQ	&	0.893	&	17.51	&	2.212139	&	1.0597E-11	&	0.552	&	39	&	M93	\\
3FGL J0315.5-1026	&	BZQ	&	1.565	&	19.71	&	2.410067	&	5.00533E-12	&		&		&		\\
3FGL J0339.5-0146	&	BZQ	&	0.852	&	18.41	&	2.252565	&	4.09816E-11	&	2.92	&	70.3	&	K10	\\
3FGL J0340.5-2119	&	BZB	&	0.223	&	17.11	&	2.217488	&	7.18461E-12	&		&		&		\\
3FGL J0405.5-1307	&	BZQ	&	0.571	&	17.09	&	2.345819	&	6.40468E-12	&	4.33	&	9.1	&	K10	\\
3FGL J0423.2-0119	&	BZQ	&	0.916	&	17	&	2.204462	&	5.95648E-11	&	2.91	&	70.2	&	K10	\\
3FGL J0424.7+0035	&	BZB	&	0.31	&	16.98	&	2.197874	&	2.1339E-11	&	1.09	&	6.1	&	K10	\\
3FGL J0428.6-3756	&	BZB	&	1.11	&	19	&	1.949819	&	1.83988E-10	&	0.68	&	86	&	CB99	\\
3FGL J0442.6-0017	&	BZQ	&	0.844	&	17	&	2.495998	&	5.12195E-11	&	2.21	&		&		\\
3FGL J0449.0+1121	&	BZQ	&	1.375	&	19.81	&	2.233381	&	4.20064E-11	&	1.56	&	15.4	&	K10	\\
3FGL J0449.4-4350	&	BZB	&	0.107	&	15.51	&	1.848965	&	1.2311E-10	&	0.0993	&	183.2	&	L08	\\
3FGL J0453.2-2808	&	BZQ	&	2.559	&	18.51	&	2.630128	&	2.49535E-11	&		&		&		\\
3FGL J0501.2-0157	&	BZQ	&	2.286	&	18.06	&	2.413701	&	3.1127E-11	&	1.66	&	148.1	&	K10	\\
3FGL J0505.3+0459	&	BZQ	&	0.954	&	18.71	&	2.463141	&	3.04786E-11	&	0.747	&		&		\\
3FGL J0530.8+1330	&	BZQ	&	2.06	&	20	&	2.245939	&	4.96196E-11	&	2.24	&	60.1	&	K10	\\
3FGL J0532.7+0732	&	BZQ	&	1.254	&	19	&	2.203017	&	6.3002E-11	&	1.54	&	126.8	&	K10	\\
3FGL J0538.8-4405	&	BZB	&	0.894	&	16.48	&	1.932384	&	3.08339E-10	&		&	220	&	CB99	\\
3FGL J0608.0-0835	&	BZQ	&	0.872	&	17.6	&	2.372678	&	2.68982E-11	&	1.2	&	123.9	&	K10	\\
3FGL J0611.1-6100	&	BZQ	&	1.773	&	20.71	&	2.704762	&	8.05141E-12	&		&		&		\\
3FGL J0630.9-2406	&	BZB	&	1.238	&	16	&	1.809364	&	4.28995E-11	&		&		&		\\
3FGL J0635.7-7517	&	BZQ	&	0.653	&	15.75	&	2.709976	&	2.04282E-11	&	0.21	&		&		\\
3FGL J0654.4+5042	&	BZQ	&	1.253	&	17.13	&	1.940565	&	1.58603E-11	&	0.31	&		&		\\
3FGL J0710.3+5908	&	BZB	&	0.125	&	15.71	&	1.661033	&	8.87552E-12	&	0.065	&	95	&	GM04	\\
3FGL J0710.5+4732	&	BZB	&	1.292	&	14.51	&	2.647554	&	9.07156E-12	&	0.973	&	94	&	M93	\\
3FGL J0721.9+7120	&	BZB	&	0.3	&	15.5	&	1.947768	&	2.12095E-10	&	0.69	&	376.4	&	K10	\\
3FGL J0733.8+5021	&	BZQ	&	0.72	&	19.3	&	2.503782	&	6.41494E-12	&	0.69	&	82.5	&	K10	\\
3FGL J0738.1+1741	&	BZB	&	0.424	&	16.22	&	2.006392	&	6.11881E-11	&	1.91	&	20.4	&	K10	\\
3FGL J0739.4+0137	&	BZQ	&	0.189	&	16.47	&	2.245392	&	3.6569E-11	&	2.34	&	40.9	&	K10	\\
3FGL J0749.0+4459	&	BZQ	&	0.192	&	16.31	&	2.371753	&	5.74985E-12	&	0.795	&	33.09	&	C04	\\
3FGL J0750.6+1232	&	BZQ	&	0.889	&	18.7	&	2.411245	&	1.25277E-11	&	1.43	&	27	&	K10	\\
3FGL J0757.0+0956	&	BZB	&	0.266	&	15	&	2.181859	&	2.11076E-11	&	2.07	&	6.7	&	K10	\\
3FGL J0808.2-0751	&	BZQ	&	1.837	&	19.8	&	1.958395	&	8.5183E-11	&	1.58	&	59.8	&	K10	\\
3FGL J0809.8+5218	&	BZB	&	0.138	&	15.21	&	1.876105	&	5.36237E-11	&	0.184	&	4.26	&	C04	\\
3FGL J0811.3+0146	&	BZB	&	1.148	&	17.2	&	2.158473	&	2.77401E-11	&	0.46	&	18.2	&	K10	\\
3FGL J0816.7+5739	&	BZB	&	0.054	&	17.41	&	2.112841	&	1.31154E-11	&	0.14	&	31.2	&	C04	\\
3FGL J0826.0+0307	&	BZB	&	0.506	&	16.8	&	2.029678	&	5.84889E-12	&	1.32	&	4.1	&	K10	\\
3FGL J0830.7+2408	&	BZQ	&	0.941	&	17.26	&	2.626917	&	1.89912E-11	&	0.76	&	62.7	&	K10	\\
3FGL J0831.9+0430	&	BZB	&	0.174	&	16.4	&	2.096903	&	3.25377E-11	&	0.8	&	150.8	&	K10	\\
3FGL J0834.1+4223	&	BZQ	&	0.249	&	17.81	&	2.439332	&	9.31559E-12	&	0.31	&		&		\\
3FGL J0841.4+7053	&	BZQ	&	2.172	&	17.3	&	2.616193	&	3.63477E-11	&	3.34	&	73.6	&	K10	\\
3FGL J0839.5+0102	&	BZQ	&	1.123	&	19.31	&	2.237636	&	6.34951E-12	&	0.349	&		&		\\
3FGL J0849.3+0458	&	BZB	&	1.069	&	18.41	&	1.969532	&	5.94958E-12	&		&		&		\\
3FGL J0854.8+2006	&	BZB	&	0.306	&	15.43	&	2.122168	&	5.79991E-11	&	1.57	&	10.7	&	K10	\\
3FGL J0903.1+4649	&	BZQ	&	1.465	&	18.66	&	2.614503	&	5.01679E-12	&	1.645	&	317	&	M93	\\
3FGL J0909.1+0121	&	BZQ	&	1.025	&	17.31	&	2.457879	&	4.40967E-11	&	1	&	38	&	K10	\\
3FGL J0909.8-0229	&	BZQ	&	0.957	&	18.81	&	2.101265	&	1.74076E-11	&		&		&		\\
3FGL J0915.8+2933	&	BZB	&	0.101	&	15.61	&	1.878973	&	2.34761E-11	&	0.222	&	111	&	A85	\\
3FGL J0912.9-2104	&	BZB	&	0.198	&	15.81	&	1.94138	&	1.15942E-11	&		&		&		\\
3FGL J0916.3+3857	&	BZQ	&	1.267	&	19.61	&	2.356156	&	3.28224E-12	&	0.62	&		&		\\
3FGL J0920.9+4442	&	BZQ	&	2.19	&	18.16	&	2.13473	&	6.14445E-11	&	1.31	&		&		\\
3FGL J0921.8+6215	&	BZQ	&	1.446	&	19.5	&	2.448463	&	1.58908E-11	&	1.11	&	6.4	&	K10	\\
3FGL J0927.9-2037	&	BZQ	&	0.347	&	16.4	&	2.383753	&	7.41932E-12	&		&		&		\\
3FGL J0929.4+5013	&	BZB	&	0.37	&	16.91	&	2.164087	&	1.15974E-11	&	0.496	&	7.94	&	C04	\\
\hline
\end{tabular}
\end{minipage}
\end{table*}
\addtocounter{table}{-1}
\begin{table*}
\centering
\begin{minipage}{105mm}
\caption{$Continue.$}
\begin{tabular}{@{}crcccccccccccccccrl@{}}
\hline\hline
3FGL name & Class & Redshift & V & $\rm{\alpha_{\gamma}}$ & $\rm{F_{\gamma} (erg s^{-1}cm^{-2})}$ & $\rm{S_{core}}$ & $\rm{S_{ext}}$ & Ref\\
{(1)} & {(2)} & {(3)} & {(4)} & {(5)} & {(6)} & {(7)} & {(8)} & {(9)}\\
\hline
3FGL J0945.9+5756	&	BZB	&	0.229	&	16.01	&	2.329607	&	7.22927E-12	&	0.069	&	9.01	&	C04	\\
3FGL J0948.6+4041	&	BZQ	&	1.249	&	18.05	&	2.668701	&	5.58658E-12	&	1.23	&	95	&	K10	\\
3FGL J0956.6+2515	&	BZQ	&	0.707	&	16.51	&	2.435683	&	1.21722E-11	&	0.483	&	19	&	M93	\\
3FGL J0957.6+5523	&	BZQ	&	0.899	&	17.89	&	1.883193	&	9.58631E-11	&	2.568	&	381	&	M93	\\
3FGL J0958.6+6534	&	BZB	&	0.368	&	16.81	&	2.38147	&	1.79248E-11	&	0.477	&	34	&	CB99	\\
3FGL J1012.7+4229	&	BZB	&	0.364	&	17.21	&	1.663461	&	4.94324E-12	&	0.07	&	10	&	C04	\\
3FGL J1015.0+4925	&	BZB	&	0.212	&	15.71	&	1.833418	&	8.86739E-11	&	0.39	&	12.31	&	C04	\\
3FGL J1018.3+3542	&	BZQ	&	1.228	&	18.17	&	2.549106	&	4.29759E-12	&	0.9	&		&		\\
3FGL J1018.8+5913	&	BZB	&	2.025	&	17.21	&	2.090416	&	3.38454E-12	&	0.074	&	10.4	&	C04	\\
3FGL J1031.2+5053	&	BZB	&	0.36	&	16.91	&	1.705211	&	1.13327E-11	&	0.038	&	1.25	&	C04	\\
3FGL J1033.2+4116	&	BZQ	&	1.117	&	18.91	&	2.321686	&	1.56188E-11	&		&		&		\\
3FGL J1037.5+5711	&	BZB	&	0.83	&	17.31	&	1.720849	&	3.43411E-11	&	0.128	&	2.18	&	C04	\\
3FGL J1053.7+4929	&	BZB	&	0.14	&	14.41	&	1.795846	&	7.04607E-12	&	0.052	&	16.6	&	C04	\\
3FGL J1051.4+3941	&	BZB	&	0.498	&	19.41	&	1.664376	&	3.25784E-12	&		&		&		\\
3FGL J1058.5+0133	&	BZB	&	0.888	&	18.28	&	2.17084	&	6.53918E-11	&	2.7	&	230.8	&	K10	\\
3FGL J1058.6+5627	&	BZB	&	0.144	&	14.61	&	1.94519	&	4.11866E-11	&	0.208	&	13.42	&	C04	\\
3FGL J1104.4+3812	&	BZB	&	0.03	&	12.8	&	1.771642	&	3.82949E-10	&	0.52	&	181	&	A85	\\
3FGL J1117.7-4632	&	BZQ	&	0.713	&	17.11	&	2.4303	&	8.24655E-12	&	0.27	&		&		\\
3FGL J1120.8+4212	&	BZB	&	0.124	&	17.51	&	1.616351	&	1.73008E-11	&	0.025	&	0.46	&	C04	\\
3FGL J1127.0-1857	&	BZQ	&	1.05	&	18.65	&	2.116971	&	7.43853E-11	&	0.66	&	12.4	&	K10	\\
3FGL J1129.9-1446	&	BZQ	&	1.184	&	16.9	&	2.786649	&	1.98472E-11	&	4.58	&	59.3	&	K10	\\
3FGL J1136.6+6736	&	BZB	&	0.134	&	15.91	&	1.719091	&	7.88996E-12	&	0.04	&	10.7	&	C04	\\
3FGL J1136.6+7009	&	BZB	&	0.046	&	11.61	&	1.823511	&	1.60471E-11	&	0.136	&	217.4	&	C04	\\
3FGL J1143.0+6123	&	BZB	&	0.475	&	17.21	&	2.02133	&	9.51823E-12	&	0.066	&	3.52	&	C04	\\
3FGL J1147.0-3811	&	BZB	&	1.048	&	16.2	&	2.253874	&	1.96677E-11	&		&	10	&	CB99	\\
3FGL J1146.8+3958	&	BZQ	&	1.089	&	19.21	&	2.320755	&	4.11165E-11	&		&		&		\\
3FGL J1150.3+2417	&	BZB	&	0.2	&	16.81	&	2.212477	&	1.59886E-11	&	0.664	&	25	&	L08	\\
3FGL J1151.4+5858	&	BZB	&	0.118	&	16.91	&	1.917273	&	9.18228E-12	&	0.137	&	55.2	&	C04	\\
3FGL J1159.5+2914	&	BZQ	&	0.725	&	14.41	&	2.094657	&	8.3383E-11	&	1.55	&	196.1	&	K10	\\
3FGL J1203.1+6029	&	BZB	&	0.066	&	12.41	&	2.208493	&	8.9288E-12	&	0.157	&	87.4	&	C04	\\
3FGL J1204.3-0708	&	BZB	&	0.184	&	15.21	&	1.861201	&	1.23818E-11	&		&		&		\\
3FGL J1205.8-2636	&	BZQ	&	0.786	&	19.5	&	2.540616	&	1.17353E-11	&		&		&		\\
3FGL J1209.4+4119	&	BZB	&	0.377	&	17.61	&	1.944873	&	5.0527E-12	&	0.397	&	1.18	&	C04	\\
3FGL J1217.8+3007	&	BZB	&	0.13	&	15.11	&	1.974478	&	6.80413E-11	&	0.355	&	189	&	A85	\\
3FGL J1221.3+3010	&	BZB	&	0.182	&	16.31	&	1.660378	&	4.66783E-11	&	0.067	&	4.3	&	GM04	\\
3FGL J1221.4+2814	&	BZB	&	0.102	&	14.91	&	2.102362	&	4.59189E-11	&	2.058	&	2.2	&	A85	\\
3FGL J1222.4+0414	&	BZQ	&	0.966	&	17.98	&	2.508516	&	2.86599E-11	&	0.6	&	155.5	&	K10	\\
3FGL J1224.9+2122	&	BZQ	&	0.432	&	17.5	&	2.185421	&	2.83853E-10	&	1.1	&	956.4	&	K10	\\
3FGL J1224.6+4332	&	BZB	&	1.075	&	20.81	&	2.63592	&	6.86639E-12	&		&		&		\\
3FGL J1229.1+0202	&	BZQ	&	0.158	&	12.85	&	2.516284	&	1.80975E-10	&	34.89	&	17671	&	K10	\\
3FGL J1231.7+2847	&	BZB	&	0.236	&	16.81	&	1.948491	&	3.09937E-11	&	0.06	&	26.3	&		\\
3FGL J1243.1+3627	&	BZB	&	1.065	&	17.01	&	1.768834	&	2.49859E-11	&	0.115	&	32.6	&	C04	\\
3FGL J1246.7-2547	&	BZQ	&	0.633	&	17.31	&	2.139547	&	1.03318E-10	&		&		&		\\
3FGL J1248.2+5820	&	BZB	&	0.847	&	15.78	&	1.947429	&	4.50237E-11	&	0.18	&	4.2	&	C04	\\
3FGL J1253.2+5300	&	BZB	&	0.445	&	17.51	&	1.897176	&	3.68671E-11	&	0.378	&	42.05	&	C04	\\
3FGL J1256.1-0547	&	BZQ	&	0.536	&	17.75	&	2.232944	&	2.39182E-10	&	10.56	&	2095	&	K10	\\
3FGL J1309.3+4304	&	BZB	&	0.69	&	17.11	&	1.935799	&	1.98727E-11	&	0.055	&	2.87	&	C04	\\
3FGL J1310.6+3222	&	BZQ	&	0.998	&	15.24	&	2.152751	&	3.74705E-11	&	1.33	&	69.1	&	K10	\\
3FGL J1317.8+3429	&	BZQ	&	1.056	&	18.21	&	2.583639	&	4.11444E-12	&	0.35	&		&		\\
3FGL J1326.8+2211	&	BZQ	&	1.4	&	18.9	&	2.333554	&	1.98206E-11	&	1.14	&	20.4	&	K10	\\
3FGL J1337.6-1257	&	BZQ	&	0.539	&	19	&	2.366261	&	1.72496E-11	&	2.07	&	151	&	K10	\\
3FGL J1351.1+0030	&	BZQ	&	2.084	&	21.81	&	2.312673	&	5.70342E-12	&		&		&		\\
3FGL J1355.0-1044	&	BZQ	&	0.332	&	18.4	&	2.386711	&	1.07919E-11	&		&		&		\\
3FGL J1417.8+2540	&	BZB	&	0.237	&	15.21	&	2.163691	&	3.9662E-12	&		&		&		\\
3FGL J1419.8+3819	&	BZQ	&	1.82	&	19.69	&	2.36463	&	7.65662E-12	&	0.52	&	2.5	&	K10	\\
3FGL J1419.9+5425	&	BZB	&	0.153	&	15.65	&	2.307488	&	1.17707E-11	&	1.058	&	18	&	A85	\\
3FGL J1427.0+2347	&	BZB	&	0.16	&	14.6	&	1.759839	&	1.48564E-10	&	0.25	&		&		\\
3FGL J1427.9-4206	&	BZQ	&	1.522	&	17.21	&	2.078681	&	1.79838E-10	&	1.37	&		&		\\
3FGL J1428.5+4240	&	BZB	&	0.129	&	15.01	&	1.574637	&	1.03283E-11	&	0.032	&	29.3	&	GM04	\\
3FGL J1434.1+4203	&	BZQ	&	1.24	&	20.21	&	2.392054	&	4.77697E-12	&		&		&		\\
3FGL J1439.2+3931	&	BZB	&	0.344	&	17.1	&	1.770915	&	5.62814E-12	&	0.038	&		&		\\
3FGL J1442.8+1200	&	BZB	&	0.163	&	15.81	&	1.795514	&	6.93295E-12	&	0.06	&	8.5	&	GM04	\\
3FGL J1454.5+5124	&	BZB	&	1.083	&	18.91	&	2.083831	&	2.68609E-11	&	0.08	&		&		\\
3FGL J1504.4+1029	&	BZQ	&	1.839	&	18.56	&	2.075091	&	2.43886E-10	&	1.82	&	38.3	&	K10	\\
3FGL J1510.9-0542	&	BZQ	&	1.185	&	17.51	&	2.425565	&	2.69772E-11	&		&		&		\\
3FGL J1512.8-0906	&	BZQ	&	0.36	&	16.54	&	2.304709	&	4.92754E-10	&	1.45	&	180.2	&	K10	\\
3FGL J1516.9+1926	&	BZB	&	1.07	&	16.91	&	2.767541	&	6.01778E-12	&	0.255	&	1.7	&	A85	\\
\hline
\end{tabular}
\end{minipage}
\end{table*}
\addtocounter{table}{-1}
\begin{table*}
\centering
\begin{minipage}{105mm}
\caption{$Continue.$}
\begin{tabular}{@{}crcccccccccccccccrl@{}}
\hline\hline
3FGL name & Class & Redshift & V & $\rm{\alpha_{\gamma}}$ & $\rm{F_{\gamma} (erg s^{-1}cm^{-2})}$ & $\rm{S_{core}}$ & $\rm{S_{ext}}$ & Ref\\
{(1)} & {(2)} & {(3)} & {(4)} & {(5)} & {(6)} & {(7)} & {(8)} & {(9)}\\
\hline
3FGL J1517.6-2422	&	BZB	&	0.049	&	13.21	&	2.112155	&	5.838E-11	&	2.562	&	32	&	A85	\\
3FGL J1540.8+1449	&	BZB	&	0.605	&	17.3	&	2.344924	&	4.65077E-12	&	1.67	&	71.4	&	K10	\\
3FGL J1542.9+6129	&	BZB	&	0.117	&	17.21	&	1.89592	&	4.48341E-11	&	0.126	&	3.7	&	C04	\\
3FGL J1549.4+0237	&	BZQ	&	0.414	&	17.45	&	2.463335	&	2.48936E-11	&	1.15	&	18.8	&	K10	\\
3FGL J1550.5+0526	&	BZQ	&	1.422	&	19.5	&	2.335909	&	1.57063E-11	&	2.21	&	42.9	&	K10	\\
3FGL J1553.5+1256	&	BZQ	&	1.29	&	17.71	&	2.223356	&	2.99178E-11	&		&		&		\\
3FGL J1558.9+5625	&	BZB	&	0.3	&	17.81	&	2.209965	&	1.06152E-11	&	0.181	&	19.7	&	C04	\\
3FGL J1607.0+1551	&	BZQ	&	0.497	&	18.11	&	2.284189	&	2.47804E-11	&		&		&		\\
3FGL J1608.6+1029	&	BZQ	&	1.226	&	18.7	&	2.619051	&	2.21968E-11	&	1.35	&	26.5	&	K10	\\
3FGL J1613.8+3410	&	BZQ	&	1.399	&	18.11	&	2.351858	&	8.24224E-12	&	2.83	&	20.6	&	K10	\\
3FGL J1635.2+3809	&	BZQ	&	1.813	&	18	&	2.414104	&	1.39949E-10	&	2.17	&	32	&	K10	\\
3FGL J1637.7+4715	&	BZQ	&	0.74	&	17.91	&	2.367525	&	2.26393E-11	&		&		&		\\
3FGL J1640.6+3945	&	BZQ	&	1.66	&	19.37	&	2.28492	&	3.56168E-11	&	1.17	&	27.6	&	K10	\\
3FGL J1653.9+3945	&	BZB	&	0.0337	&	13.78	&	1.716347	&	1.29752E-10	&	1.376	&	67	&	A85	\\
3FGL J1700.1+6829	&	BZQ	&	0.301	&	19.01	&	2.398081	&	3.93633E-11	&	0.313	&		&		\\
3FGL J1719.2+1744	&	BZB	&	0.137	&	19.1	&	2.042681	&	1.75946E-11	&	0.661	&	11	&	A85	\\
3FGL J1725.3+5853	&	BZB	&	0.297	&	17.1	&	2.178923	&	4.96372E-12	&	0.052	&	20.4	&	C04	\\
3FGL J1728.5+0428	&	BZQ	&	0.296	&	18.31	&	2.594238	&	1.48769E-11	&	0.98	&		&		\\
3FGL J1727.1+4531	&	BZQ	&	0.717	&	18.1	&	2.346577	&	1.99875E-11	&	1	&	55.3	&	K10	\\
3FGL J1728.3+5013	&	BZB	&	0.055	&	14.51	&	1.95986	&	1.22073E-11	&	0.175	&	50	&	A85	\\
3FGL J1730.6+3711	&	BZB	&	0.204	&	17.1	&	2.08777	&	5.04089E-12	&	0.062	&	40.09	&	C04	\\
3FGL J1733.0-1305	&	BZQ	&	0.902	&	19.5	&	2.244674	&	6.15628E-11	&	6.13	&	517.8	&	K10	\\
3FGL J1740.3+5211	&	BZQ	&	1.375	&	18.7	&	2.452514	&	2.32386E-11	&	1.61	&	27.6	&	K10	\\
3FGL J1743.9+1934	&	BZB	&	0.084	&	13.31	&	1.776926	&	7.95062E-12	&	0.157	&		&		\\
3FGL J1742.2+5947	&	BZB	&	0.4	&	17.31	&	2.301161	&	5.32817E-12	&	0.106	&	5	&	C04	\\
3FGL J1749.1+4322	&	BZB	&	0.215	&	18.31	&	2.24691	&	1.3187E-11	&	0.235	&	19.5	&	C04	\\
3FGL J1751.5+0939	&	BZB	&	0.322	&	16.78	&	2.118364	&	3.8702E-11	&	1.05	&	4.9	&	K10	\\
3FGL J1748.6+7005	&	BZB	&	0.77	&	16.21	&	2.064401	&	4.59656E-11	&	0.61	&	12	&	CB99	\\
3FGL J1801.5+4403	&	BZQ	&	0.663	&	17.9	&	2.695124	&	9.26074E-12	&	0.5	&	246.6	&	K10	\\
3FGL J1800.5+7827	&	BZB	&	0.684	&	15.9	&	2.222699	&	5.86176E-11	&	1.98	&	20.8	&	K10	\\
3FGL J1806.7+6949	&	BZB	&	0.051	&	14.22	&	2.234023	&	3.89695E-11	&	1.2	&	368.7	&	K10	\\
3FGL J1813.6+3143	&	BZB	&	0.117	&	15.81	&	1.910873	&	1.12677E-11	&	0.074	&		&		\\
3FGL J1824.2+5649	&	BZB	&	0.664	&	19.3	&	2.459963	&	2.82805E-11	&	0.95	&	137.4	&	K10	\\
3FGL J1829.4+5402	&	BZB	&	0.177	&	17.81	&	1.931645	&	7.22787E-12	&	0.018	&		&		\\
3FGL J1838.8+4802	&	BZB	&	0.3	&	16.81	&	1.806124	&	1.44346E-11	&	0.051	&	1.2	&	C04	\\
3FGL J1849.2+6705	&	BZQ	&	0.657	&	16.9	&	2.142674	&	4.92003E-11	&	0.47	&	101	&	K10	\\
3FGL J2001.0-1750	&	BZQ	&	0.65	&	18.6	&	2.309653	&	2.23784E-11	&	1.82	&	9.4	&	K10	\\
3FGL J2000.0+6509	&	BZB	&	0.047	&	12.51	&	1.883226	&	6.81896E-11	&	0.2	&	60	&	GM04	\\
3FGL J2005.2+7752	&	BZB	&	0.342	&	16.7	&	2.218685	&	2.06807E-11	&	0.823	&	28.9	&	M93	\\
3FGL J2025.6-0736	&	BZQ	&	1.388	&	19.31	&	2.181554	&	7.29303E-11	&		&		&		\\
3FGL J2035.3+1055	&	BZQ	&	0.601	&	16.37	&	2.467874	&	2.50098E-11	&	0.781	&	40	&	A85	\\
3FGL J2056.2-4714	&	BZQ	&	1.489	&	18.1	&	2.2647	&	7.73787E-11	&		&		&		\\
3FGL J2134.1-0152	&	BZB	&	1.285	&	19	&	2.207499	&	1.28413E-11	&	1.37	&	151.9	&	K10	\\
3FGL J2143.5+1744	&	BZQ	&	0.211	&	15.73	&	2.403724	&	6.39508E-11	&	0.386	&		&		\\
3FGL J2147.2+0929	&	BZQ	&	1.113	&	18.54	&	2.377022	&	3.56426E-11	&	0.698	&	82	&	M93	\\
3FGL J2158.0-1501	&	BZQ	&	0.672	&	18.3	&	2.26945	&	1.26202E-11	&	2.7	&	304.7	&	K10	\\
3FGL J2158.8-3013	&	BZB	&	0.117	&	13.41	&	1.750328	&	2.30501E-10	&	0.252	&	132	&	A85	\\
3FGL J2202.7+4217	&	BZB	&	0.069	&	14.72	&	2.161141	&	1.77287E-10	&	1.99	&	14.2	&	K10	\\
3FGL J2203.4+1725	&	BZQ	&	1.075	&	19.5	&	2.152252	&	4.93254E-11	&	0.87	&	74.6	&	K10	\\
3FGL J2212.0+2355	&	BZQ	&	1.125	&	20.66	&	2.212319	&	1.26704E-11	&	0.43	&	0.9	&	K10	\\
3FGL J2217.0+2421	&	BZB	&	0.505	&	18.51	&	2.215108	&	1.09927E-11	&	0.42	&		&		\\
3FGL J2225.8-0454	&	BZQ	&	1.404	&	18.39	&	2.358605	&	2.47089E-11	&	7.13	&	91.6	&	K10	\\
3FGL J2229.7-0833	&	BZQ	&	1.56	&	17.43	&	2.388687	&	5.40756E-11	&	0.93	&	8.4	&	K10	\\
3FGL J2232.5+1143	&	BZQ	&	1.037	&	17.33	&	2.339162	&	7.26124E-11	&	6.99	&	148	&	K10	\\
3FGL J2236.3+2829	&	BZQ	&	0.795	&	19.01	&	2.115377	&	4.35777E-11	&	1.118	&	3.4	&	M93	\\
3FGL J2243.4-2541	&	BZB	&	0.774	&	17.81	&	2.274767	&	1.95586E-11	&		&		&		\\
3FGL J2254.0+1608	&	BZQ	&	0.859	&	16.1	&	1.634632	&	1.23418E-09	&	14.09	&	822	&	K10	\\
3FGL J2258.0-2759	&	BZQ	&	0.927	&	16.77	&	2.173132	&	4.87036E-11	&	2.58	&		&		\\
3FGL J2258.3-5526	&	BZB	&	0.479	&	18.51	&	2.310879	&	3.16722E-12	&		&		&		\\
3FGL J2319.2-4207	&	BZB	&	0.054	&	14.81	&	2.095091	&	4.76094E-12	&	0.25	&	391.9	&	L08	\\
3FGL J2334.1+0732	&	BZQ	&	0.401	&	16.04	&	2.504762	&	1.01509E-11	&	0.61	&	38.4	&	K10	\\
3FGL J2338.1-0229	&	BZQ	&	1.072	&	19.01	&	2.495937	&	1.85497E-11	&		&		&		\\
3FGL J2348.0-1630	&	BZQ	&	0.576	&	18.41	&	2.202846	&	2.93969E-11	&	1.99	&	142.7	&	K10	\\
3FGL J2359.3-3038	&	BZB	&	0.165	&	16.41	&	2.02159	&	6.61874E-12	&	0.039	&	27.2	&	GM04	\\
\hline
\end{tabular}
\medskip{The $\rm{\gamma}$-ray fluxes and photon spectral indexs come from the 3FGL (Fermi-LAT Collaboration,2015).}
\end{minipage}
\end{table*}

\clearpage
\clearpage
\begin{table*}
\centering
\begin{minipage}{100mm}
\caption{Correlation analysis\label{tbl-2}}
\begin{tabular}{@{}crrrrrrrrrrr@{}}
\hline\hline
Param 1 & Param 2 & Class & Pearson & Pearson \\
        &         &       &   Coeff &   prob  \\
(1)& (2) & (3) & (4) & (5)\\
\hline
$\rm{L_{\gamma}}$	&	$\rm{L_{core}}$	&	All	&	0.921	&	$\rm{1.84\times10^{-68}}$	\\
$\rm{L_{\gamma}}$	&	$\rm{L_{ext}}$ 	&	All	&	0.855	&	$\rm{7.65\times10^{-42}}$	\\
$\rm{L_{\gamma}}$	&	$\rm{L_{core},z^{\ast}}$	&	All	&	0.766	&	$\rm{7.13\times10^{-33}}$	\\
$\rm{L_{\gamma}}$	&	$\rm{L_{ext},z^{\ast}}$ 	&	All	&	0.644	&	$\rm{7.31\times10^{-18}}$	\\
$\rm{L_{\gamma}}$	&	$\rm{R_{c}}$	&	All	&	0.176	&	0.041	\\
$\rm{L_{\gamma}}$	&	$\rm{R_{v}}$	&	All	&	0.719	&	$\rm{1.41\times10^{-27}}$	\\
$\rm{L_{\gamma}}$	&	$\rm{[\log L_{ext}]/M_{abs}}$	&	All	&	0.233	&	0.005	\\
$\rm{L_{\gamma}}$	&	$\rm{[\log L_{ext}]/M_{abs},z^{\ast}}$	&  All	&	0.078	&	0.356	\\
                    &                                   &   FSRQs &  0.230  &   0.064   \\
$\rm{L_{\gamma}}$	&	$\rm{[\log L_{ext}]/M_{abs},z^{\ast}}$	&  FSRQs	&	-0.032	&	0.801	\\
                    &                                   &   BL Lacs & 0.401 &   $\rm{2.5\times10^{-4}}$ \\
$\rm{L_{\gamma}}$	&	$\rm{[\log L_{ext}]/M_{abs},z^{\ast}}$	&  BL Lacs	&	0.24	&	0.034	\\
$\rm{L_{\gamma}}$	&	$\rm{M_{abs}}$	&	All	&	-0.715	&	$\rm{9.55\times10^{-33}}$	\\
$\rm{\alpha_{\gamma}}$	&	$\rm{z}$	&	FSRQs	&	0.016	&	$\rm{0.869}$	\\
                        &               &	BL Lac	&	0.324	&	$\rm{0.001}$	\\

\hline
\end{tabular}
\medskip{columns (1) and (2): parameters being examined for correlations.\\ Results from a Pearson analysis are cited in the main text. ``$\ast$'' is Pearson\\ partial correlation analysis excluding redshift. significant correlation P$<$0.05.}
\end{minipage}
\end{table*}
\end{document}